\documentclass[aps,prl,twocolumn,showpacs,floatfix,superscriptaddress]{revtex4}
\usepackage{graphicx}
\usepackage{amsmath}
\usepackage{amssymb}
\bibstyle{apsrev.bib}

\usepackage{subfigure} 

\usepackage{epsfig}

\begin{document}

\title{Self-Organisation in 2D Swarms.}
 
\author{Jihad Touma}
\affiliation{Center for Advanced Mathematical Studies (CAMS) and Physics Department, 
American University of Beirut, Beirut, Lebanon.} 
\email{jt00@aub.edu.lb}

 \author{Amer Shreim} 
\affiliation{Physics Department, 
American University of Beirut, Beirut, Lebanon.}
\email{amer.shreim@gmail.com}

\author{Leonid I. Klushin}
\affiliation{American University of Beirut, Department of Physics, 
Beirut, Lebanon}

\date{\today}

\begin{abstract}
  We undertake a systematic numerical exploration of self-organised
  states in a deterministic model of interacting, self-propelled
  particles in 2D. In the process, we identify new types of collective
  motion, namely disordered swarms, rings and droplets. We construct a
  ``phase diagram'', which summarizes our results as it delineates
  phase transitions (all discontinuous) between disordered swarms and
  vortical flocks on one hand, and bound vortical flocks and expanding
  formations on the other. One of transition lines is found to have a 
  close analogy in the temperature-driven gas-liquid transition in finite 
  clusters with the same interparticle potential. 
  Furthermore, we report on a novel type of
  flocking which takes place in the presence of a uniform external
  driver.  Altogether, our results set a rather firm stage for
  experimental refinement and/or falsification of this class of
  models.
\end{abstract}

\pacs{
05.65.+b  
}

\maketitle 

Populations of self-propelled organisms tend to organize in remarkable
aggregate formations~\cite{parrish}. Schools of fish, flocks of birds,
swarms of bees or locust are among the more familiar examples. Less
familiar perhaps, though equally prevalent, are the organized states
achieved by microscopic organisms: amoeba aggregates
\cite{levineAmoeba}, bacterial colonies~\cite{vicsekBacteria,
  bacterialrings}, and swarms of Daphnia~\cite{daphnia}. Such striking
self-organized collective phenomena stimulated natural extensions of
trusted models and tools of equilibrium statistical mechanics to
systems of coupled self propelled particles. Proposed models have
varied in degrees of idealization and complexity of the description of
swarms, their environments and the interactions between them. They can
be broadly divided into continuous~\cite{mogilner, flierl,
  tu&toner1995, tu&toner1998} and discrete self-propelled particle
(SPP) models~\cite{vicsek1995, vicsek2000, levine, shimoyama}, both
coming in deterministic as well as stochastic varieties. In these
pioneering studies, models were mainly analyzed for their ability to
display certain observed phenomena (transition from swarm to flock
\cite{vicsek1995, tu&toner1995}, vortical swarms~\cite{levine,
  ModelDaphnia1}), as well as the potential universality of certain
features (scaling behavior~\cite{vicsek2000}, individual distance
\cite{keshet}). More often than not, they were not pushed to
predictive risks that would at best qualify and refine them, at worst
falsify them.  While these exercises are crucial to identify generic
sufficient conditions for the occurrence of observed states, they
leave open the question as to whether or not these conditions
necessarily obtain in the living system under investigation. In cases
where such systems are open to detailed experimental investigation
\cite{bacterialrings}, it proves useful to explore the richness of
behavior sustained by classes of models, with a view to proposing
tracks for spatio-temporal evolution that might result under (slow)
changes in the experimental setting.  This is precisely the objective
we had in mind as we embarked on our explorations, and shall dedicate the
rest of this letter to describing the working model (discrete,
deterministic, SPP model), summarizing our extensive numerical
experiments, and highlighting our significant, and eminently testable,
results: a- novel types of self-organisation namely the \emph{disk,
  ring, polarized vortex, droplet} and \emph{expansion} states; b- a
phase diagram which distills, on a reduced parameter plane, the
various states, and the transitions between them; c- an intriguing
route to flocking, which obtains in the presence of an external,
uniform, force field.

Given the inherent complexity of our undertaking, we opted for a
simple, but fairly versatile, deterministic SPP model, flavors of
which have been examined by Levine \emph{et. al.}~\cite{levine}, in
their quest for vortices, and Edelstein-Keshet \emph{et.al}
\cite{keshet} in a study of individual distances in swarms. The
particles in this model are identical in their mass, and in the nature
of the forces they feel and generate. Their self-propulsion is
mimicked with an acceleration of constant magnitude, acting along the
direction of motion. They are coupled via a double exponential
potential force field, which attracts at large distances, and repels
at small distances. Furthermore, they are subjected to a drag
force, which captures the reaction of the viscous medium in which they
move (and which is here assumed linear in the velocity).  Newton's
equation for the $i$th particle reads
\begin{equation}
\label{equationsofmotion1}
m \frac{d\vec{v_i}}{d t} = \sigma \hat{v_i} - \gamma \vec{v_i} - \nabla_{\vec{r_i}} \phi 
\end{equation}
\noindent
where $m$, $\vec{r_i}$ and $\vec{v_i}$ are the mass, position and
velocity of the $i$th particle, $\sigma$ the magnitude of the
self-propulsion force, which acts along the direction of motion
$\hat{v_i}$, and $\gamma$ the friction coefficient.  The potential energy
$\phi$ is given by
\begin{equation}
\label{potential}
\phi = \sum_{\stackrel{i, j = 1}{j \neq i}}^{N} W_r exp(-|\vec{r_i} - \vec{r_j}|/l_r) - W_a exp(-|\vec{r_i} - \vec{r_j}|/l_a)
\end{equation}
\noindent
where $N$ is the population size, and $W_r$, $W_a$, $l_r$ and $l_a$
determine the strength and range of repulsive and attractive forces
respectively. It proves useful to work with the dimensionless
variables $\vec{r'} = \vec{r}/l_r$, $\vec{v'} = \vec{v}/v_t$, and $t'
= t/\tau$, with $v_t=\frac{\sigma}{\gamma}$ (the ``terminal
velocity''), and $\tau =l_r/v_t = l_r \gamma/\sigma$. In the primed
variables, the equation of motion becomes
\begin{equation}
\label{dimensionlessequationsofmotion1}
R \frac{d \vec{v}'_i}{d t'} = \hat{v}'_i - \vec{v}'_i  +
  \sum_{\stackrel{j=1}{j \neq i}}^{N} \left[ Q exp(-|\vec{r}'_i - \vec{r}'_j|) - P
  exp(-|\vec{r}'_i - \vec{r}'_j|/{\lambda}) \right ] \hat{r}_{ij}'
\end{equation}
\noindent
with $\hat{r}_{ij}' = [\vec{r}'_j - \vec{r}'_i]/|\vec{r}'_j -
\vec{r}'_i|$. We recover four dimensionless parameters:
$\lambda = l_r/l_a$ (repulsive over attractive length scale, $< 1$),
$P = W_a/\sigma l_a$ (attraction over self-propulsion) , $Q=
W_r/\sigma l_r$ (repulsion over self-propulsion), and $R = m
\sigma/(l_r {\gamma}^2)$ (a Reynolds-like number for the flow), which
together with the population size $N$, fully specify a swarm in our
model.

In what follows, we are primarily concerned with the behavior of 2D
swarms. Initially (and except when otherwise specified), particles are
uniformly distributed in a square (its size being chosen so that the
initial swarm is bounded~\cite{note1}), with randomly oriented initial
velocities, and speeds uniformly distributed between 0 and the
terminal velocity. Populations ranging in size between 100 and 1000
members were considered. With these initial conditions, and the help
of a 4th order, adaptive, Runge-kutta scheme~\cite{numericalrecipes},
we thoroughly explored the parameter space. Simulations were allowed
to run till the system's energy (and wherever relevant, mean velocity
and density) relaxed to near steady configurations, on which our
classification is based. Results are viewed in the center of mass
frame, the origin of which, though interesting in its wanderings, is
not a serious concern of this inquiry.

\begin{figure}[!tp]
\begin{center}
\epsfxsize= 6.3 cm
\epsfysize= 14 cm
\epsfbox{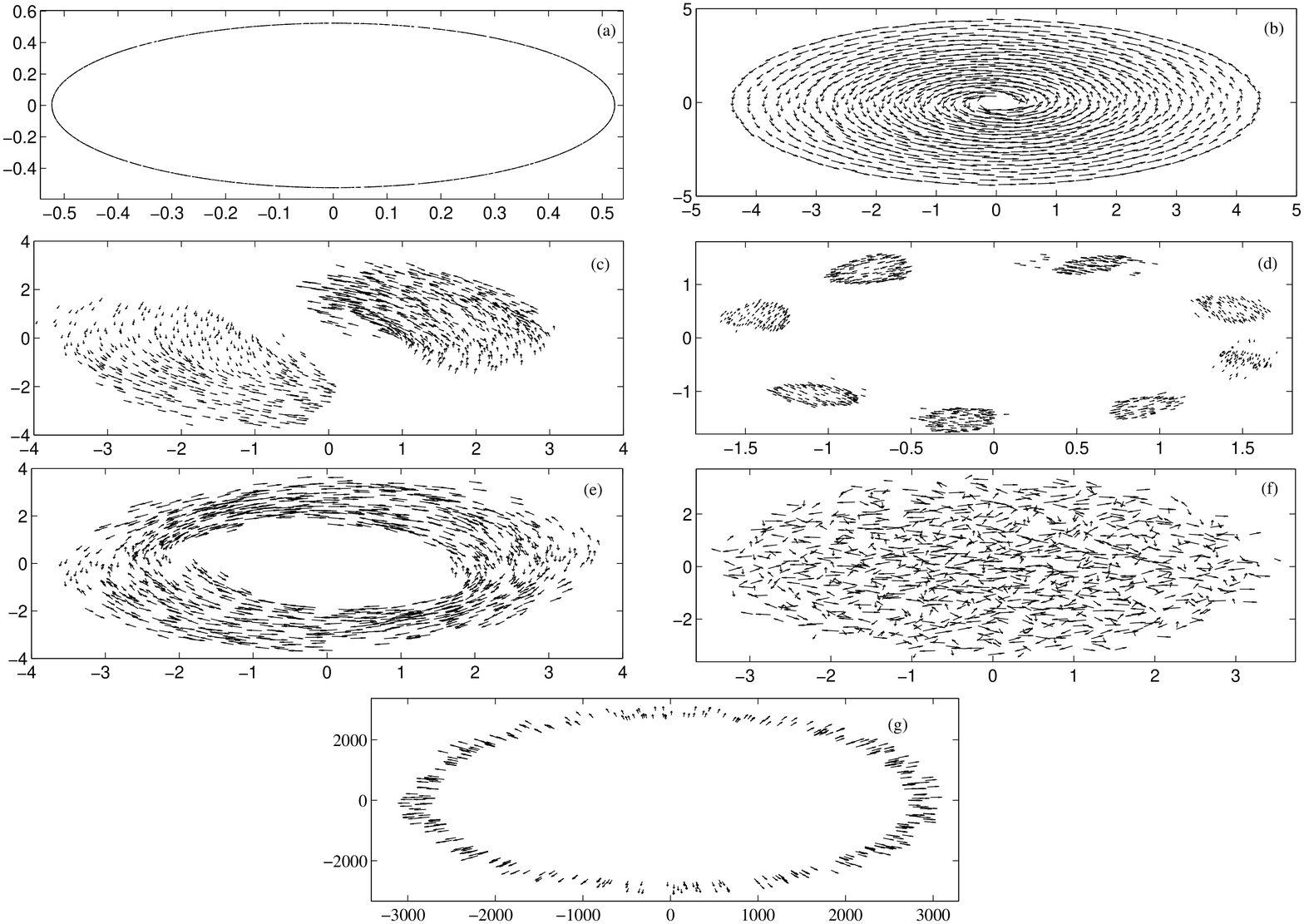}
\end{center}
\caption[regimes]{\label{ring} Steady states, for $N=1000$, $\lambda =
  2/3$, and $P=0.1$: (a) Ring State, $Q=0.1$ and $R=3$. (b)
  Polarized vortex, $Q=0.21$ and $R=0.0975$. (c) and (d) Droplet states,
  with $Q=0.1575, R=3.75$ and $Q=0.1125, R=7.5$ respectively. (e) Vortex
  state, $Q=0.15$ and $R=5.625$.  (f) Disordered Disk State, $Q=0.1875$ and
  $R=0.0375$; (g) Expansion State, $Q=0.2625$ and $R=0.375$.}
\label{regimes}
\end{figure}

Broadly speaking, these random initial conditions converge, to a
center manifold (think of it as the terminal velocity manifold), on
which the swarm relaxes to distinct regimes, which though not
exhaustive of the rich behavior allowed by this model, constitute the
skeleton on which a complete description will be eventually fleshed
out. Before displaying, then analyzing, a typical (two dimensional)
space in which these regimes find there proper home, we survey their
main qualitative properties. We draw our results from experiments with
1000 particles, while noting that the existence and main qualitative
properties of these regimes are independent of the size of the
population.

We start with the \emph{ring} state, shown in Fig.~\ref{regimes}(a), a
highly regimented configuration, in which particles rotate and
counter-rotate on a perfect circle. Next in line, is the
\emph{polarized vortex}, shown in Fig.~\ref{regimes}(b), in which
nearly all particles travel in the same direction~\cite{note2}. A hitherto unsuspected formation is the
\emph{droplet} state, Fig.~\ref{regimes}(c-d), in which the swarm
breaks up into a necklace of drop-like flocks. Droplets rotate
clockwise and counter-clockwise, with near constant angular velocity,
holding tight through repeated mergers with fellow droplets traveling
in the opposite direction. Bona fide \emph{vortices}, on which particles
are nearly split even between prograde and retrograde circulation, are
shown in Fig.~\ref{regimes}(e).  Fig.~\ref{regimes}(f) shows an example
of the disordered \emph{disk} state, in which particles move
chaotically, while self-confined within a disk like region. Lastly,
the \emph{expansion} state, in which self-propulsion dominates over
attractive forces to promote the outward explosion of the swarm, is
shown in Fig.~\ref{regimes}(g).
 
\begin{figure}[!ht]
\begin{center}
\epsfxsize= 8.5  cm
\epsfysize= 6 cm
\epsfbox{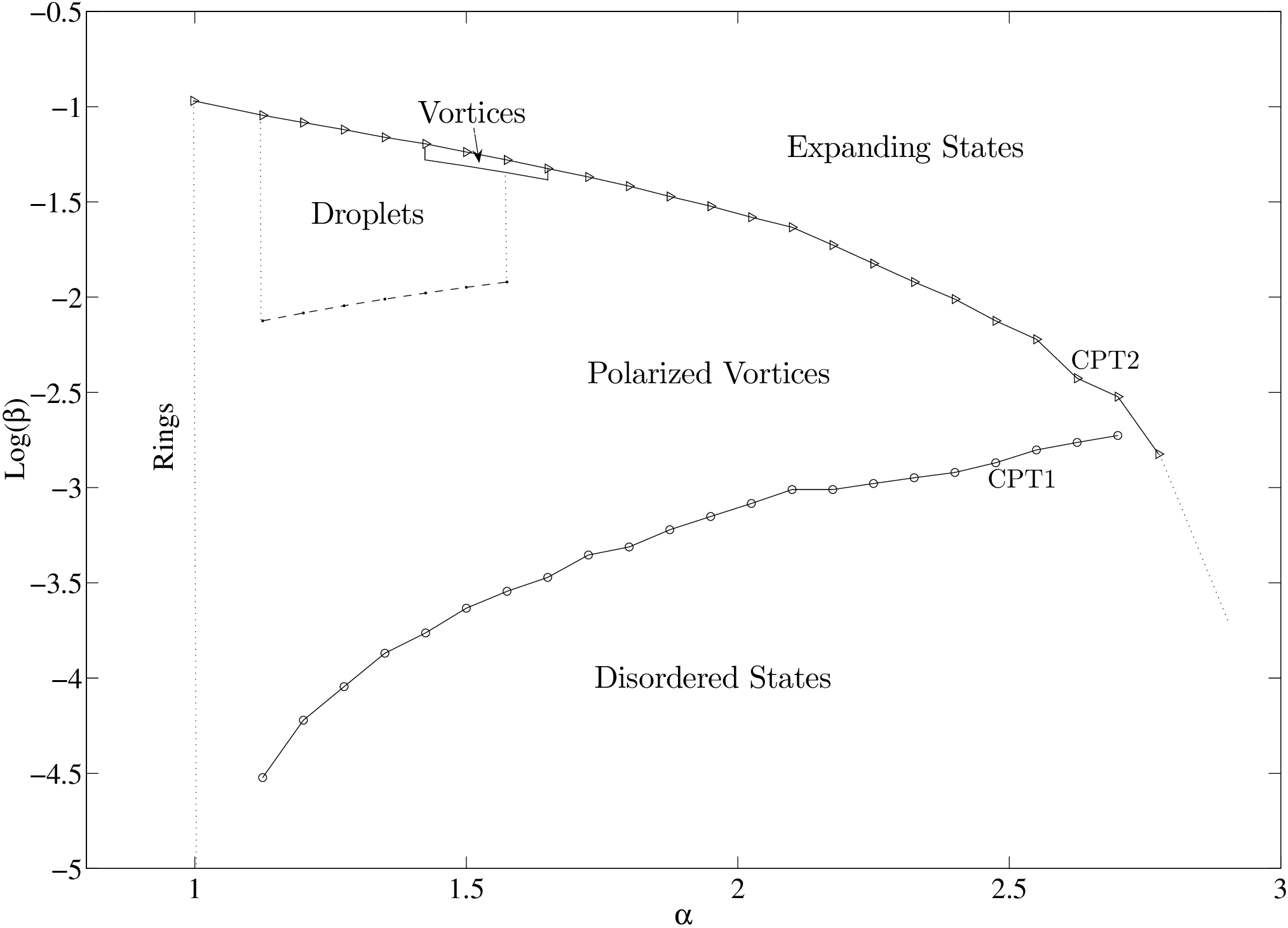}
\end{center}
\caption{\label{PhaseDiagram} The phase diagram, for $N=1000$, and
  $\lambda = 2/3$. $\beta = R/NP$ is plotted against $\alpha = Q/P$.
  The dotted $\alpha = 1$ line, the line of \emph{rings}, separates
  the multi-swarm (attraction dominated) region to its left, from the
  swarm dynamics regions to its right. Circles mark transitions from
  disordered \emph{disk} to \emph{polarized vortex} state; triangles
  mark transitions from ordered to \emph{expansion} state.}
\end{figure}

Perhaps more important than the identification of novel self-organized
states, is the realization that the five dimensional parameter space
(the four dimensionless model parameters, plus $N$) reduces, once
the dynamics settles on the terminal velocity manifold, to a three
dimensional space, in which the variety of relaxed, self-organized
configurations, is itself organized in a phase (morphology) diagram of
sorts. This space is spanned by $\lambda = l_r/l_a$, $\alpha = Q/P =
W_r/(\lambda W_a)$, and $\beta = R/NP = m v_t^2/(N \lambda W_a)$. We
verified, with extensive numerical explorations, that models with the
same ($\lambda$, $\alpha$, $\beta$) relax to qualitatively similar
configurations, that a steady configuration in the $(\lambda, \alpha,
\beta)$ plane unfolds into a hyper-surface of similar configurations in
the five dimensional parameter space of the model. 

It was noted~\cite{bertozzi} that the Morse potential (Eq.~\ref{potential}) with parameter values resulting in self-organized states is not $H$-stable in the language of statistical mechanics~\cite{ruelle}. $H$-stability is related to the existence of a well-defined thermodynamic limit with finite intensive local densities. The system which is not $H$-stable is called catastrophic. Indeed, the necessary criterion to ensure this stability is that the integral over space of the pair interaction potential be non-negative. Since
\[
\int {\phi (r)d^2 r = 2\pi W_a l_a ^2 \left( {\alpha \lambda ^3  - 1} \right)} 
\]
the criterion is not satisfied and the system is catastrophic for $\alpha  < \lambda ^{ - 3}$. In the catastrophic regime the average potential energy per particle in large but finite systems is proportional to $N$. 

The reduced set of parameters has, therefore, simple physical meaning: $\alpha$ and $\lambda$ together define the shape of the potential, and, in particular, the degree to which the system is away from $H$-stability. As $\alpha$ increases towards the value of $\lambda ^{ - 3}$ the system approaches the normal, non-catastrophic regime. On the other hand,  $\beta$  has the meaning of the ratio of the typical kinetic energy to the potential energy per particle (which is dominated by attraction and is of order $N W_a$). 

In this reduced parameter space, one could survey steady states, and
neighbouring configurations, by varying three of the model parameters
(say $W_a$, $l_a$ and $\gamma$ or $\sigma$) and keeping the others
($W_r$, $l_r$, $N$ and $\sigma$ or $\gamma$) fixed.  Thousands of
computing hours went into clarifying the salient features of the
resulting phase diagram, the invariant structures of the model. These
features are clearly delineated on constant $\lambda$ slices, where
the relaxed dynamics typically splits into four regions, as apparent
in the particular instance in Fig.~\ref{PhaseDiagram}: a) a region
occupied by the disordered \emph{disk} states, for smallish $\beta$
(attraction and viscosity dominated models); b) a region occupied by unbound,
expanding states, for largish $\beta$ (propulsion dominated models);
c) an intermediate region where all organized states live (i.e. rings,
droplets, polarized vortices and regular vortices). This region is
bounded below by the critical phase transition line (CPT1) between
disordered and ordered states, and above by the critical phase
transition line (CPT2) between ordered and expanding states. It is
bounded to the left by the line $\alpha = 1$, or $W_a/l_a =W_b/l_b$, a
line of ring states, the radius of which increases with increasing
$\beta$, all else being held fixed, till the point (past CPT2) where
the dominant self-propulsion fragments the ring into an expanding
state; d) a region of irregular, multi-cluster, and unsteady
configurations, which is obtained in attraction dominated models (to
the left of the \emph{ring} line, and which we keep out of our humble
focus on coherent, relaxed swarms).

\begin{figure}[!ht]
\begin{center}
\epsfxsize= 7.5 cm
\epsfysize= 4 cm
\epsfbox{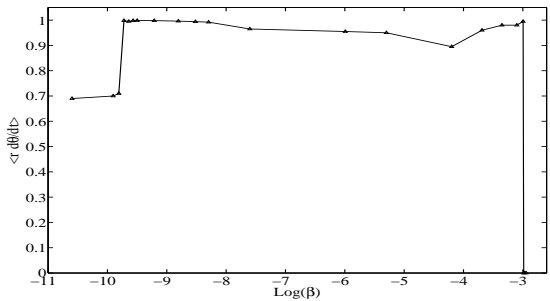}
\end{center}
\caption{\label{PhaseTransition} The transition among CPT1 and CPT2
are shown. The order parameter $<r|\dot{\theta}|>$ is plotted against $\beta$.
The graph was obtained for $N=1000$, $\lambda=2/3$, and $\alpha = 1.275$.}
\end{figure}

CPT1 and CPT2 meet at a crossroads between the disordered disk, the
polarized vortex and the expanding state; CPT2 stretches beyond that
meeting point into regimes which are held together by viscosity's
countering of the mutual repulsion of particles. For $\alpha > 1$, and
$\beta$ just above CPT1, polarized vortices are always observed. In
fact, just above CPT1, \emph{all} particles are circulating in the
same direction. Increasing $\beta$, at constant $\alpha$, the number of counter
rotating particles grows steadily till the polarized vortex is
destabilized, in three possible ways, depending on the value of
$\alpha$: 1) The polarized vortex breaks up into droplets (mostly two)
rotating in the same direction, with a small fraction of particles
moving erratically in the opposite direction. Increasing $\beta$, the
number of droplets increases, with nearly as many droplets rotating in
one direction as the other. Looking at larger values of $\beta$, the
droplets can either go directly to the expansion state or, and for a
small $\alpha$ interval, they become more elongated and pass through
the vortex state before breaking up and expanding. 2) The polarized
vortex morphs directly into a regular vortex and then to an expanding
state. 3) The polarized vortex transitions directly to the expansion
state; here, and very close to CPT2, we find mixed states--that is
states in which a small fraction of particles expands while the other
forms a special vortex~\cite{note3}.  Transitions among the organized
states are harder to identify numerically, better handled through a
careful stability analysis in a mean field approach to the problem,
and are only roughly determined in the current study~\cite{note4}. On
the other hand, transitions across CPT1 and CPT2 are discontinuous,
and reminiscent of the first order phase transitions of equilibrium
statistical mechanics. To better clarify this character, we define an
order parameter $L = <r|\dot{\theta}|>$, the bracket indicating an
average over all the particles in a given steady state.  $L$ is nearly
equal to zero in the expanding state, tends to $\frac{2}{\pi} \approx
0.636$\ for a completely random state, and $1$ for vortex like
dynamics.  Following $L$ with $\beta$, for fixed $\alpha$, we observe
the system in Fig.~\ref{PhaseTransition} undergoing the two sharp
transitions, from random to organized, and from organized to
expanding. For $\alpha > 2.7$, the swarm transitions directly from the
\emph{disordered} to the \emph{expansion} state.

To clarify the nature of some of the transitions observed, we undertake a study of a thermodynamic analogy of the self-propelling system. 

It is clear from the phase diagram that one of the control parameters, $\beta$,  has the meaning of the ratio of the typical kinetic energy to the potential energy per particle. In the thermodynamic analogy,we consider a system with exactly the same interaction, but the kinetic energy is controlled by temperature instead of the self-propulsion to viscosity ratio. Consider the canonical ensemble description of a system of $N$ particles with pair interaction specified by  Eq.~\ref{potential} enclosed in a circular box of radius $R$ and kept at temperature $T$. Thermodynamic properties are treated in the self-consistent field approximation.
According to the standard formulation of the variational mean field theory~\cite{chaikin} the density is related to the mean-field potential by

\[
n(\vec r) = A\exp \left( {\frac{{ - \varphi _{mf} (\vec r)}}{T}} \right)
\]

where A is determined by normalization 
\[
N = \int {n(\vec r)d^2 r}
\]

and the mean- field potential is given by
\[
\varphi _{mf} (\vec r) = \int {n(\vec r')\phi(\vec r' - \vec r)d^2 r'} 
\]
Axial symmetry was assumed, and self-consistent field equations were solved numerically by an iterative procedure for various temperatures and interaction parameter $\alpha$ at fixed $\lambda$=2/3.
In a fairly broad temperature range two stable solutions exist: one with an almost uniform low density, corresponding to the gas phase, and the other with a large density near the center and extremely low density at the periphery, corresponding to the liquid droplet phase.
Intriguingly, stable droplet solutions exist only within the catastrophic regime $\alpha  < (3/2)^3$.

In order to analyze thermodynamic stability we calculate the Helmholtz free energy for each solution (phase) in the same self-consistent field approximation:
\[
\begin{array}{l}
F = \frac{1}{2}\int {n(r)n(r')\phi (|\vec r - \vec r'|) d^2 r d^2 r'} +\\
\\
\quad \quad + T\int {n(r)\ln \left( {n(r)/N} \right) d^2 r}\\
 \end{array}
\]

\begin{figure}
\begin{center}
\includegraphics[scale=0.3]{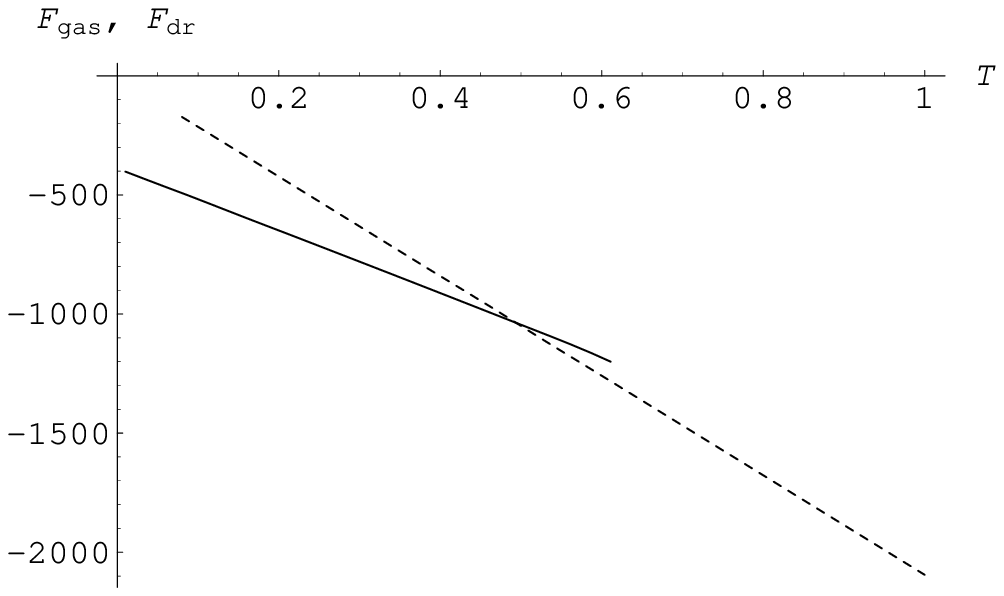}
\hspace{0.4cm}
\includegraphics[scale=0.3]{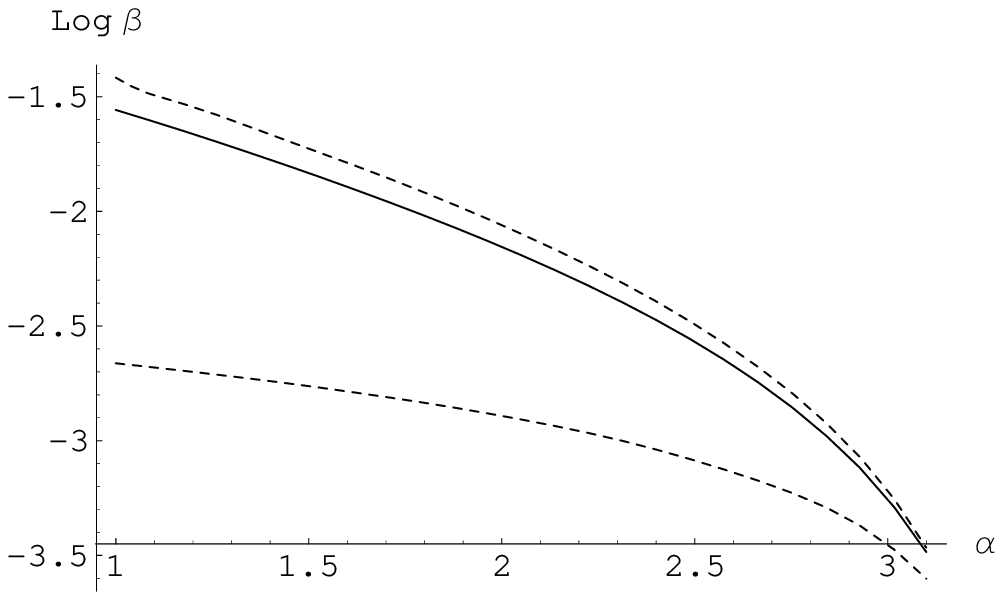}
\caption{\label{thermo} (a) Temperature dependence of the two branches of the free energy: liquid droplet phase (solid line) and gas phase (dotted line) for $N=1000$, $\lambda=2/3$, $\alpha=2.85$. The lines terminate at respective spinodal points.$T$ is expressed in $W_a$ units. 
(b)Phase diagram of the thermodynamic system for $N=1000$, $\lambda=2/3$, box radius $R=30 l_a$, in the same coordinates as in Fig~\ref{PhaseDiagram}. The equilibrium transition line is solid, spinodal lines are dashed. The liquid droplet phase is stable below the transition line, and metastable in the range between the solid and the upper dashed lines.}
\end{center}
\end{figure}

The velocity-related contribution is the same for both phases and will not be considered explicitly.
The solution with the lower free energy corresponds to thermodynamic equilibrium while the other solution can be associated with a metastable phase.
Upon approaching certain temperature, a solution may lose it iterative stability and eventually disappear. In the language of thermodynamics, this would mean reaching the spinodal line. Temperature dependence of the two branches of the free energy is shown in Fig.~\ref{thermo}(a).

The fact that two branches of the free energy cross is typical of classical first order transitions. However, in contrast to standard situations simultaneous coexistence of two phases is impossible. Each of the two solutions describes the system as a whole while, say, a 50-50 mixture of uniform gas and liquid droplet is not a self-consistent solution at all. This is why the transition temperature is defined from the condition that the two phases have equal total free energies rather than from equality of chemical potentials. We recall that another distinction of the current situation is that a standard thermodynamic limit is not very meaningful. However, from a pragmatic point of view, first order transitions in finite systems can still be defined if there are two distinct states with the free energies that are large on the $T$ scale and cross as functions of some control parameter. Phase transitions in an isolated macromolecule are conceptually close to the situation studied here,~\cite{Escape}

We make a connection with the dynamic system by equating $m {v_t}^2$ and $T$ (Boltzmann constant is 1) since the system is 2-dimensional, so that $\beta=T/N\lambda W_a$ . The transition line in the ($\alpha$, Log$\beta$) coordinates is presented in Fig.~\ref{thermo}(b) together with two spinodal lines for $N$=1000, $\lambda$=2/3, and $R=30 l_a$. Separate calculations with a different number of particles $N$=500 confirm that our choice of $\beta$ and $\alpha$ as scaling parameters yield a universal $N$-independent phase diagram. There is a weak (logarithmic) effect of the box radius $R$ on the transition temperature which is naturally related to the entropy of the gas phase, but further elaboration of the thermodynamic model is the subject of future work.

It is clear that the gas-droplet transition line resembles closely the CPT2 line separating the expanding and the ordered states. This gives a strong support to the claim that the transitions in the dynamic system are indeed a direct analogy to phase transitions. Analysis of the dynamic analogy of the metastable thermodynamic states would involve a study of the attractor basins in the phase space and is beyond the scope of the present paper. The thermodynamic system does not support dynamically organized states, nor do we observe a transition analogous to the CPT1 line. A system of self-propelled particles can be highly organized in the momentum space which is impossible for a classical system of interacting particles in the canonical ensemble where Maxwell's distribution is inavoidable. Note, however, that the current statements concerning the thermodynamic system are based only on the results of the self-consistent field approximation, while the true ordering in the coordinate space may be more intricate.

\begin{figure}[!ht]
\begin{center}
\epsfxsize= 8.5 cm
\epsfysize= 3.5 cm
\epsfbox{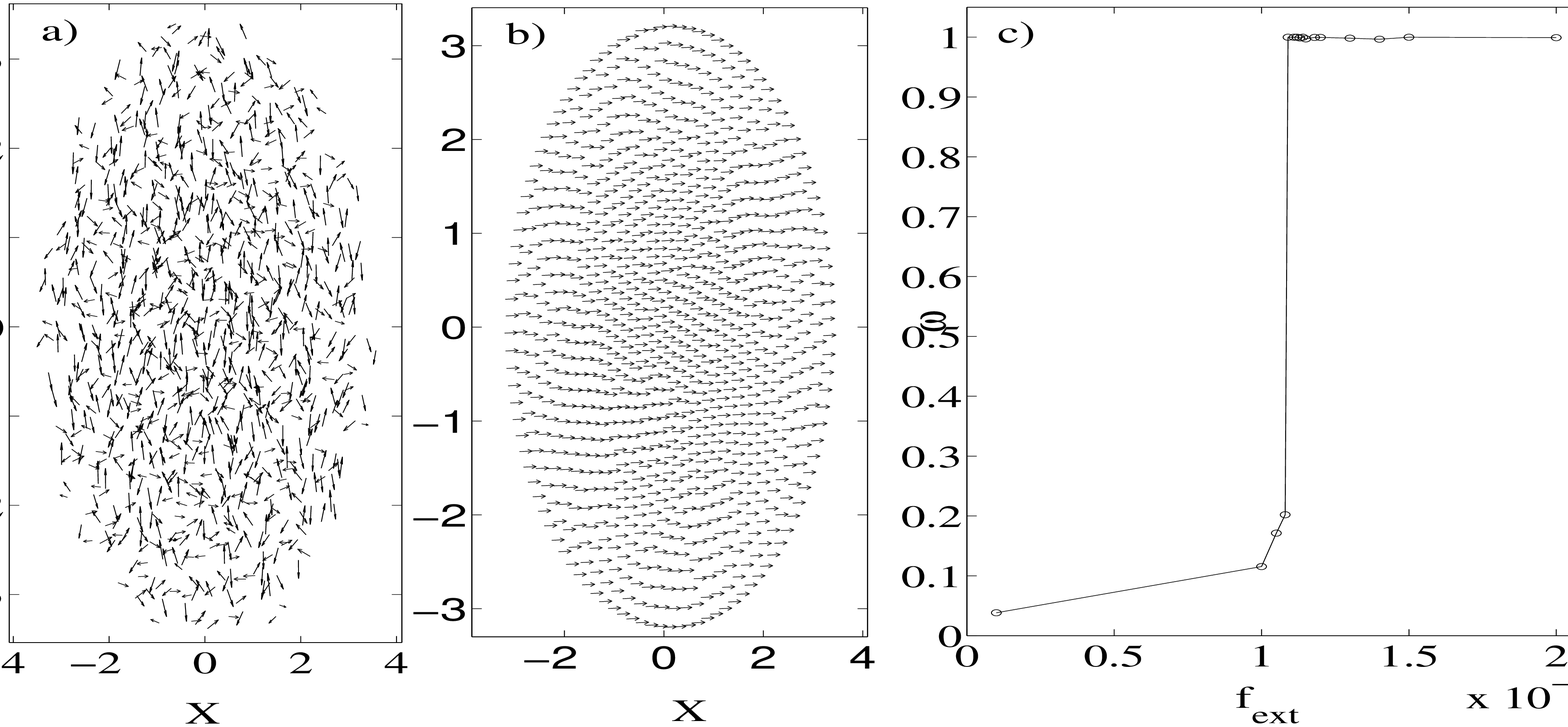}
\end{center}
\caption[]{\label{SwarmInExtFieldPhaseTrans} The transition from disordered
swarm to an aligned flock: a) Shows the initial disordered 
state; b) Shows the steady solution after applying the external field ($f_{ext} =
0.00115$); c) shows the 1st order phase transition where we plot the order parameter
$\omega$ vs the external field strength (normalized by self-propulsion
$\sigma$). The graphs were obtained for $N=1000$, $\lambda = 2/3$, $\alpha=1.875$ and $\beta = 5.25\times10^{-5}$.}
\end{figure}

We conclude our discussion of critical phase transitions with a brief
mention of an interesting phenomenon, which is analogous to the
notorious swarm to flock transition, and which, unlike the
self-promoted transitions seen in this and other instances in the
literature, occurs in the presence of an external, uniform, force
field. Fig.~\ref{SwarmInExtFieldPhaseTrans} (b) shows a
\emph{polarized} flock in which all particles are traveling with the
same velocity, a state which resulted by subjecting a
\emph{disordered} disk state of our model (seen in
Fig.~\ref{SwarmInExtFieldPhaseTrans} (a)) to a constant, external force
field. Such a polarized flock does not emerge for any field strength.
In fact, a discontinuous transition, from random to polarized, occurs
as the strength of the external field, ${\vec{f}}_{ext}$, is increased
past a critical value (which naturally depends on the initial phase of
the swarm (i.e. on the swarm's $\lambda, \alpha$, and $\beta$). The
center of mass speed: $\omega = | \sum_{i = 1}^{N} \vec{v_i}|/(N
v_{max})$ (normalized by $v_{max}$, the maximum speed in the swarm), is a
suitable order parameter for the transition from \emph{disordered}
disk ($\omega=0$), to \emph{polarized} flock ($\omega = 1$), which is 
clearly illustrated in Fig.~\ref{SwarmInExtFieldPhaseTrans}(c).

In summary, we have succeeded in constructing a phase diagram that
captures, in one fell swoop, the global structure of steady 2D
configurations, in a deterministic model of coupled self-propelled
particles. Through a systematic numerical exploration of the model's
parameter space, we have identified novel types of collective behavior
such as \emph{rings}, \emph{droplets}, \emph{special vortices},
\emph{disordered} disks, and \emph{expanding} states; we have shown
that transitions from disordered to ordered, then from ordered to
expanding state, have the structure of discontinuous, ``first order'',
phase transitions (in line with the results of~\cite{gregoire}, which
attribute the continuous transitions reported by~\cite{vicsek1995} to
finite-size effects); we have supported the phase transition analysis 
by studying a thermodynamic analogy where a temperature-driven transition 
from a gas to a condensed liquid droplet is observed and matches 
closely the transition from ordered to expanded state; finally, we have 
shown that a \emph{disordered}
swarm, when subjected to an external uniform field, transitions to a
\emph{polarized} flock, once again in a discontinuous manner. The
genericity of discontinuous phase transitions, the stability and
evolution of organized states, the surface tension and shape dynamics
of swarms, are the subject of ongoing work. In the mean time, the
constructed phase diagram offers a promising geography for
experimental verification, and refinement, or at worst, falsification
of this class of models. In particular, the same bacteria which
stimulated an earlier investigation of model vortices~\cite{levine},
could perhaps be stimulated into transitions from disordered motion,
to polarized vortex, to droplets, or vortex, by judicious selection
and control of population size, medium resistance, concentrations of
chemo-attractants/repellents, external drivers...etc. As evident in
Fig.~\ref{PhaseDiagram}, an order of magnitude increase in population
size (decrease in $\beta$) is expected to bring about the
disintegration of a colony, initially organized in a polarized vortex,
into a disordered disk, and this over a range of interaction
potentials (of $\alpha$s); similar effects could result from a slowing
down of particles in the colony (through increased friction and/or
reduced self-propulsion); an increasing concentration of
chemo-attractants (a decreasing $\alpha$) may push an initially
vortical colony closer to a \emph{ring} state~\cite{note5}. We look
forward to exchanges with experimentalists around this model's phase
diagram, in the hope that similar systematic (not to say exhaustive)
explorations of macroscopic phases of collective motion in the lab,
together with the model improvements that they will surely stimulate,
may eventually pave the way to a refined characterization of the
notoriously elusive (at times microscopic) mechanical properties
of coupled, self-propelled living organisms.

\bibliography{SwarmReferences}

\begin{thebibliography}{28}
\bibitem[{not({\natexlab{c}})}]{note3}
\bibinfo{note}{All the states refered to here are steady states solutions
  reached by starting from random initially conditions, and not by following a
  relaxed state with slowly changing parameters.}
\end{thebibliography}
 
\end{document}